\documentclass[conference]{IEEEtran}
\IEEEoverridecommandlockouts
\usepackage{cite}
\usepackage{amsmath,amssymb,amsfonts}
\usepackage{algorithmic}
\usepackage{graphicx}
\usepackage{textcomp}
\usepackage{xcolor}
\usepackage{multirow}
\usepackage{mathtools}

\usepackage{subfigure}

\usepackage{comment}

\usepackage{xspace}

\newcommand{\etal}{\textit{et al.}}

\def\BibTeX{{\rm B\kern-.05em{\sc i\kern-.025em b}\kern-.08em
    T\kern-.1667em\lower.7ex\hbox{E}\kern-.125emX}}
    
\begin{document}

\title{Where to Move Next: Zero-shot Generalization of LLMs for Next POI Recommendation}








\author{{Shanshan Feng${^{1,2}}\scriptsize^{\dag}$, Haoming Lyu${^{3}}\scriptsize^{\dag}$, Fan Li${^{4}}^*$, Zhu Sun${^{1,2}}$, Caishun Chen${^{1,2}}$}
 \\
	\fontsize{9}{9}\selectfont\itshape
	
    $^{1}$Centre for Frontier AI Research, A*STAR, Singapore;\\
    $^{2}$Institute of High Performance Computing, A*STAR, Singapore;\\
    $^{3}$Nanyang Technological University, Singapore; \\
	$^{4}$Hong Kong Polytechnic University, Hong Kong, China \\
	\fontsize{8}{8}\selectfont\ttfamily\upshape
 
    victor\_fengss@foxmail.com; haominglyu@gmail.com;fan-5.li@polyu.edu.hk;\\
    sunzhuntu@gmail.com;chen\_caishun@cfar.a-star.edu.sg	

\thanks{
($\dag$Both authors contributed equally to this research;$^*$Corresponding author.)
}

}

\maketitle

\begin{abstract}
Next Point-of-interest (POI) recommendation provides valuable suggestions for users to explore their surrounding environment. Existing studies rely on building recommendation models from large-scale users' check-in data, which is task-specific and needs extensive computational resources. Recently, the pretrained large language models (LLMs) have achieved significant advancements in various NLP tasks and have also been investigated for recommendation scenarios. However, the generalization abilities of LLMs still are unexplored to address the next POI recommendations, where users' geographical movement patterns should be extracted. Although there are studies that leverage LLMs for next-item recommendations, they fail to consider the geographical influence and sequential transitions. Hence, they cannot effectively solve the next POI recommendation task. To this end, we design novel prompting strategies and conduct empirical studies to assess the capability of LLMs, e.g., ChatGPT, for predicting a user's next check-in. Specifically, we consider several essential factors in human movement behaviors, including user geographical preference, spatial distance, and sequential transitions, and formulate the recommendation task as a ranking problem. Through extensive experiments on two widely used real-world datasets, we derive several key findings. Empirical evaluations demonstrate that LLMs have promising zero-shot recommendation abilities and can provide accurate and reasonable predictions. We also reveal that LLMs cannot accurately comprehend geographical context information and are sensitive to the order of presentation of candidate POIs, which shows the limitations of LLMs and necessitates further research on robust human mobility reasoning mechanisms. 
\end{abstract}


\section{Introduction} \label{sec:intro}
Recent years have witnessed the rapid development of location-based social networks (LBSNs) such as Foursquare and Facebook Places, where users can share their geographical positions by checking in points of interest (POI) on social networks. POIs usually denote the specific geographical locations that some users might find useful or interesting, such as coffee shops and libraries. Based on the check-in records, we can learn the user's mobility movement patterns and further recommend appropriate POIs for users to visit. The POI recommendation task~\cite{CSUR_survey_sanchez2022point} is of great value in real-world scenarios, as it can help users to better explore their surroundings, attract potential consumers for business holders, and increase the revenue of service platforms.

\begin{figure}[!ht]
\vspace{-1ex}
\centering
\includegraphics[width=2.8in]{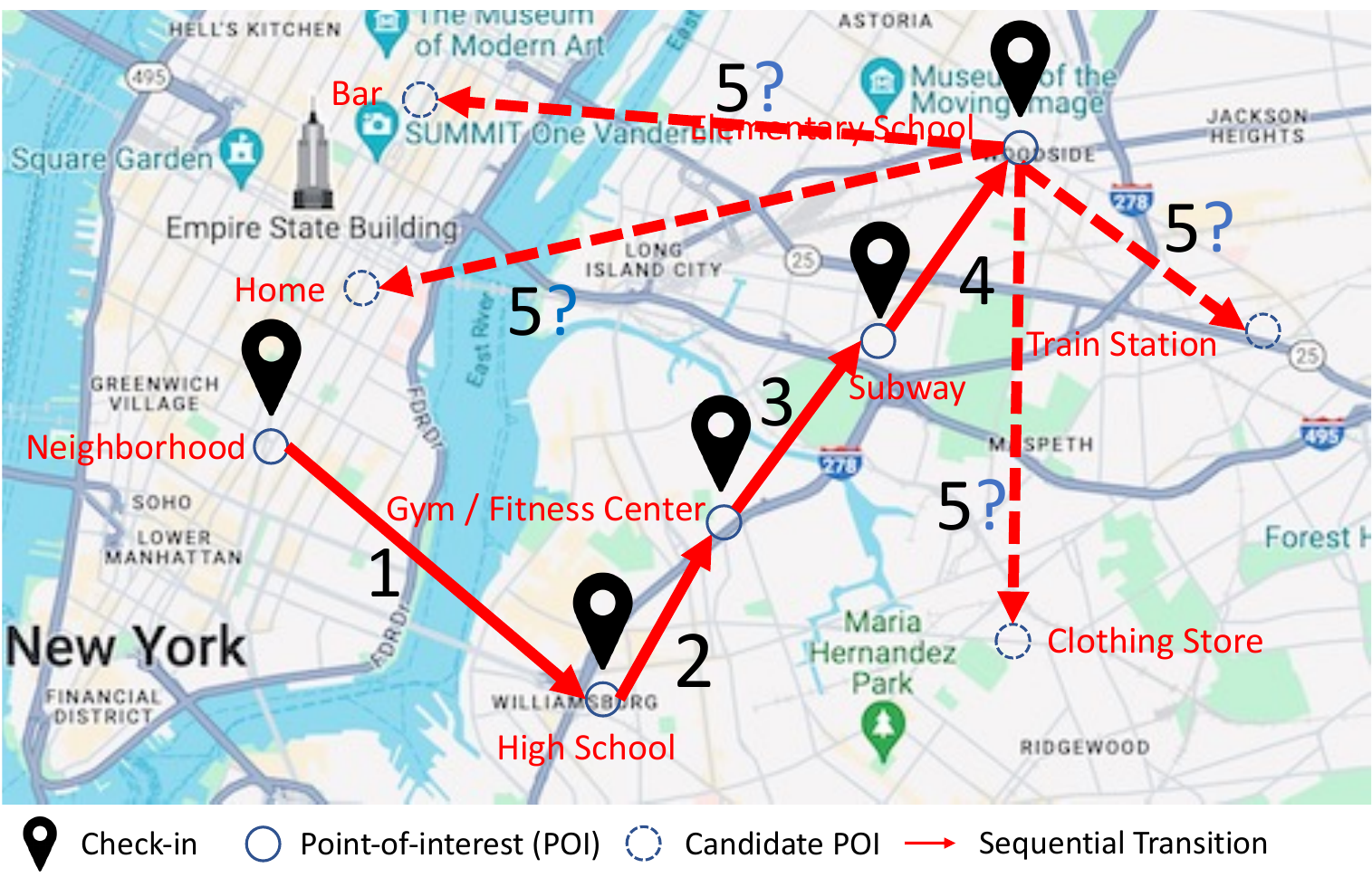}
\vspace{-1ex}
\caption{An example of the next POI recommendation task. Each POI is associated with geographical coordinates and category information. Given the recent trajectory, e.g., $\{l_1, l_2, l_3, l_4\}$, it aims to predict POIs to visit subsequently.} \label{fig:introduction}
\vspace{-1ex}
\end{figure}

Compared with conventional POI recommendation tasks, the next POI recommendation task~\cite{IJCAI15_feng2015personalized} focuses specifically on predicting the user's next likely visit, which is more challenging. As presented in Figure~\ref{fig:introduction}, given a user's check-in trajectory $\{l_1, l_2, l_3, l_4\}$, it aims to recommend the next location to visit. 
The next POI recommendation problem has attracted extensive research interest and various recommendation models have been developed~\cite{feng2017poi2vec,AAAI23_yin2023next, SIGIR22_yang2022getnext,SIGIR20_feng2020hme,CIKM23_duan2023clsprec,yan2023_st_hypergraph_sigir}. However, existing methods need construct and train recommendation models from large-scale users' check-in data, which consumes extensive computational resources. In addition, these recommendation methods are task-specific and lack generalization capability. In this work, we examine the next POI recommendation task from another perspective. Instead of training a task-specific recommendation model, we attempt to leverage the general-purpose pretrained large language models for generating sequential POI suggestions, which has not been explored before.

LLMs not only have achieved remarkable results in various natural language processing tasks, but also have shown impressive performance in many domains. First, the geographical information can be extracted from the pretrained LLMs and further used for spatial-temporal studies. For example, Manvi \etal~\cite{GEOLLM_standord_manvi2023geollm} prove that LLMs embed remarkable spatial information, and Gurnee \etal~\cite{LLM_ST_MIT_gurnee2023language} find that LLMs learn linear representations of space and time across multiple scales. Roberts \etal~\cite{GPT4GEO_roberts2023gpt4geo} examine the degree to which GPT-4 acquires factual geographical knowledge and interpretative reasoning. Second, LLMs have been used for analyzing user mobility behaviors and spatial trajectory patterns~\cite{Arxiv23_MOB_wang2023would, Arxiv_23_Anomaly_zhang2023large, Arxiv_23_Public_events_liang2023exploring}. However, these preliminary studies do not fully consider the geographical correlations and focus on different targets, e.g., anomaly trajectory detection and public event prediction,  which are dissimilar to our work. Last, several LLM-based next-item methods~\cite{wang2023zero_SMU, chatgpt_zero_liu2023chatgpt, Zero-shot-ranker-hou2023large, list_ranking_dai2023uncovering} have been proposed and obtained promising zero-shot sequential recommendation performance on movie and E-commerce dataset, etc. Current solutions, unfortunately, fall short of capturing crucial aspects for next-Point of Interest (POI) recommendation tasks, specifically the geographical correlations and sequential transitions. Consequently, there is a pressing need to explore effective strategies for leveraging LLMs to address personalized user mobility recommendations. 

This study investigates the application of pretrained LLMs in modeling human check-in trajectory data. Utilizing LLMs for this purpose presents a non-trivial challenge, as LLMs are inherently designed and optimized for language processing, making direct usage impractical for location prediction. To overcome this limitation, we introduce a novel framework named LLMmove, aiming to seamlessly integrate human movement prediction with language modeling. One of the key differences between next POI recommendation and the next-item recommendation~\cite{wang2023zero_SMU, chatgpt_zero_liu2023chatgpt, Zero-shot-ranker-hou2023large, list_ranking_dai2023uncovering} is the geographical correlations in user movements since users tend to visit close locations rather than far away places. This assumption is consistent with Tobler's first law of geography, ``Everything is related to everything else, but near things are more related than distant things", which is the fundamental assumption used in spatial analysis. Specifically, we present the check-in data into \emph{long-term check-ins} and \emph{recent check-ins}, which reflects the user's long-term spatial preference and the current spatial preference, respectively. For each POI in the \emph{candidate set}, we calculate its geographical distance\footnote{We also explore many strategies to use ChatGpt to directly compute the spatial distances based on the POIs' coordinates, but cannot obtain accurate results. Hence, we calculate the distances and utilize them as input.} from the user's current position (indicated by the last check-in). Here, we incorporate four important factors for the next POI recommendation: long-term preference, current preference, geospatial distance, and potential sequential transitions. Then, by considering these requirements, LLMs are instructed to recommend Top-K POIs and provide explanations for the returned recommendations. 

We conduct extensive experiments on two widely used real-world datasets for next POI recommender systems, which yield several significant insights. Our empirical evaluations showcase the promising zero-shot recommendation capabilities of LLMs, providing relatively accurate and reasonable predictions. However, it becomes evident that LLMs struggle with accurately grasping geographical context information and exhibit sensitivity to the order in which candidate POIs are presented. These limitations underscore the need for further research to develop robust human mobility reasoning mechanisms in conjunction with LLMs.

The contributions of this work are summarized as follows:
\begin{itemize}
    \item We investigate a novel research task, which explores the zero-shot generalization of LLMs to address the next POI recommendation. To the best of our knowledge, this is the first work to utilize the LLMs for POI recommendations. 
    \item We develop a novel prompting framework, namely LLMmove, to incorporate various factors for sequential POI recommendation, including user spatial preferences, geographical distances, and sequential transitions.
    \item We conduct extensive experiments on two real-world datasets and derive several findings. The empirical results demonstrate the effectiveness of the proposed framework. The datasets and codes are available at https://github.com/LLMMove/LLMMove.
\end{itemize}
\section{Related Work}

\subsection{Next-POI Recommendation}
As an important human mobility mining task, the next-POI recommendation problem captures the users' complex personalized check-in behaviors, where various factors play essential roles including individual interests, continuous movement patterns, and spatial-temporal influence, etc. 
Recently, the next POI recommendation has attracted extensive research interests and a large variety of approaches have been developed ~\cite{IJCAI22_zhang2022next,AAAI23_yin2023next, SIGIR20_feng2020hme,SIGIR22_yang2022getnext,TKDD_ou2023sta,CIKM23_duan2023clsprec,sun2023_TOIS,yan2023_st_hypergraph_sigir}. 
However, existing approaches require constructing and training recommendation models using extensive users' check-in data, demanding significant computational resources. Moreover, these task-specific recommendation models fall short in providing zero-shot POI recommendations for users. Different from them, this work aims to generate the next POI suggestions without the need for task-specific training, a direction not previously explored.

\subsection{LLM-Based Recommender Systems}
Very recently, the LLMs have been exploited for the recommendation tasks~\cite{fan2023_recsys_in_era_LLM}. 
Although LLMs are not specifically designed for capturing user-item interactions, their proficiency in understanding textual information and robust generative capabilities, including providing explanations and justifications, holds significant promise for improving recommendations. An illustrative example is the generative GPT4Rec framework proposed in~\cite{li2023gpt4rec}, which treats the recommendation task as a query generation and searching procedure.

Several LLM-based recommendation methods address sequential recommendation problems. Harte \etal~\cite{Recsys_23_harte2023leveraging} propose three variants: LLM Embeddings, Fine-Tuned LLM, and LLM-enhanced Sequential Model. Wang \etal~\cite{wang2023zero_SMU} introduce Zero-Shot Next-Item Recommendation with a prompting strategy guiding GPT-3 through user preferences, historical items, and top-K recommendations. Liu \etal~\cite{chatgpt_zero_liu2023chatgpt} evaluate ChatGPT in five recommendation scenarios, employing zero-shot and few-shot prompt strategies for next-item prediction based on past sequential behaviors. \cite{Zero-shot-ranker-hou2023large} enhances sequential recommendations with a recency-focused prompting method. Dai \etal~\cite{list_ranking_dai2023uncovering} combine ChatGPT with information retrieval for improved recommendation capabilities. However, these prompt-based methods lack consideration for geographical information, hindering their effectiveness in solving the next POI recommendation task.

\subsection{LLMs for User Mobility Patterns}

The exploration of leveraging pre-trained models for modeling geographical spatial data has garnered increasing research attention. Two main approaches emerge: training geospatial pre-trained models and utilizing open-accessible Large Language Models (LLMs) for geospatial tasks. Mai \etal~\cite{GEOAI_mai2023opportunities} and Balsebre \etal~\cite{CityFM_balsebre2023cityfm} focus on geospatial foundation models. Based on the open-sourced LLaMA model, Deng \etal~\cite{WSDM24_deng2023K2} develop a foundation language model for understanding and utilizing geoscience knowledge. For open-accessible LLMs, studies like \cite{GeoGPT_zhang2023geogpt,LLM-GEO_li2023autonomous,GPT4GEO_roberts2023gpt4geo} explore tasks like population description, economic livelihood measurement, and route planning. \cite{GEOLLM_standord_manvi2023geollm} and \cite{LLM_ST_MIT_gurnee2023language} highlight that LLMs capture spatial information and acquire coherent knowledge about space and time.

Approaches like \cite{Arxiv_23_Anomaly_zhang2023large, Arxiv_23_Public_events_liang2023exploring, Arxiv23_MOB_wang2023would} employ pre-trained LLMs for various human mobility prediction tasks, such as anomaly detection using LLMs~\cite{Arxiv_23_Anomaly_zhang2023large}, predicting travel demand under public events~\cite{Arxiv_23_Public_events_liang2023exploring}. 
It's worth mentioning that \cite{Arxiv23_MOB_wang2023would} introduces the LLM-Mob framework, utilizing accessible LLMs for learning mobility data. While it accounts for both long-term and short-term dependencies, its primary focus is on incorporating temporal information into human mobility sequences. However, the framework is specifically designed for the time-aware location prediction task, limiting recommendations to historically visited places. In essence, it does not provide recommendations for new locations and does not consider the geographical information of places.
\section{Problem Statement} \label{sec:def}


Let $L = \{l_1, l_2, \ldots, l_{|L|}\}$ represent a set of Points of Interest (POIs), with each POI denoted as $<Id, Cat, Lat, Lon>$. Here, $Id$ denotes the unique ID for the respective POI, $Cat$ indicates its category (e.g., Gym or Train station) providing semantic information, and $Lat$ and $Lon$ signify geographical coordinates, specifying latitude and longitude, respectively. Each check-in is represented as a tuple $c^u_{l,t} = <u, l, t>$, indicating that user $u$ visited POI $l$ at timestamp $t$. A trajectory $traj=\{c^u_{l_1,t_1}, c^u_{l_2,t_2}, \ldots, c^u_{l_k,t_k}\}$ represents a sequence of POIs visited by a user within a short time interval (e.g., 24 hours in this study). Focusing primarily on POI sequences in this work, the trajectory is denoted as $traj=\{l_1, l_2, \ldots, l_k\}$ to avoid ambiguity.

Building upon prior research~\cite{SIGIR22_yang2022getnext,yan2023_st_hypergraph_sigir}, the objective of next POI recommendation is to furnish a list of potential POIs that a user is likely to visit subsequently. Formally, given the historical check-ins of a specific user and their current trajectory $traj=\{l_1, l_2, \ldots, l_k\}$, the aim is to predict the probable next POI $l_{k+1}$ that will be visited in the near future.


\section{Methodology} \label{sec:solu}

\begin{figure}[!ht]
\vspace{-2ex}
\centering
\includegraphics[width=3.3in]{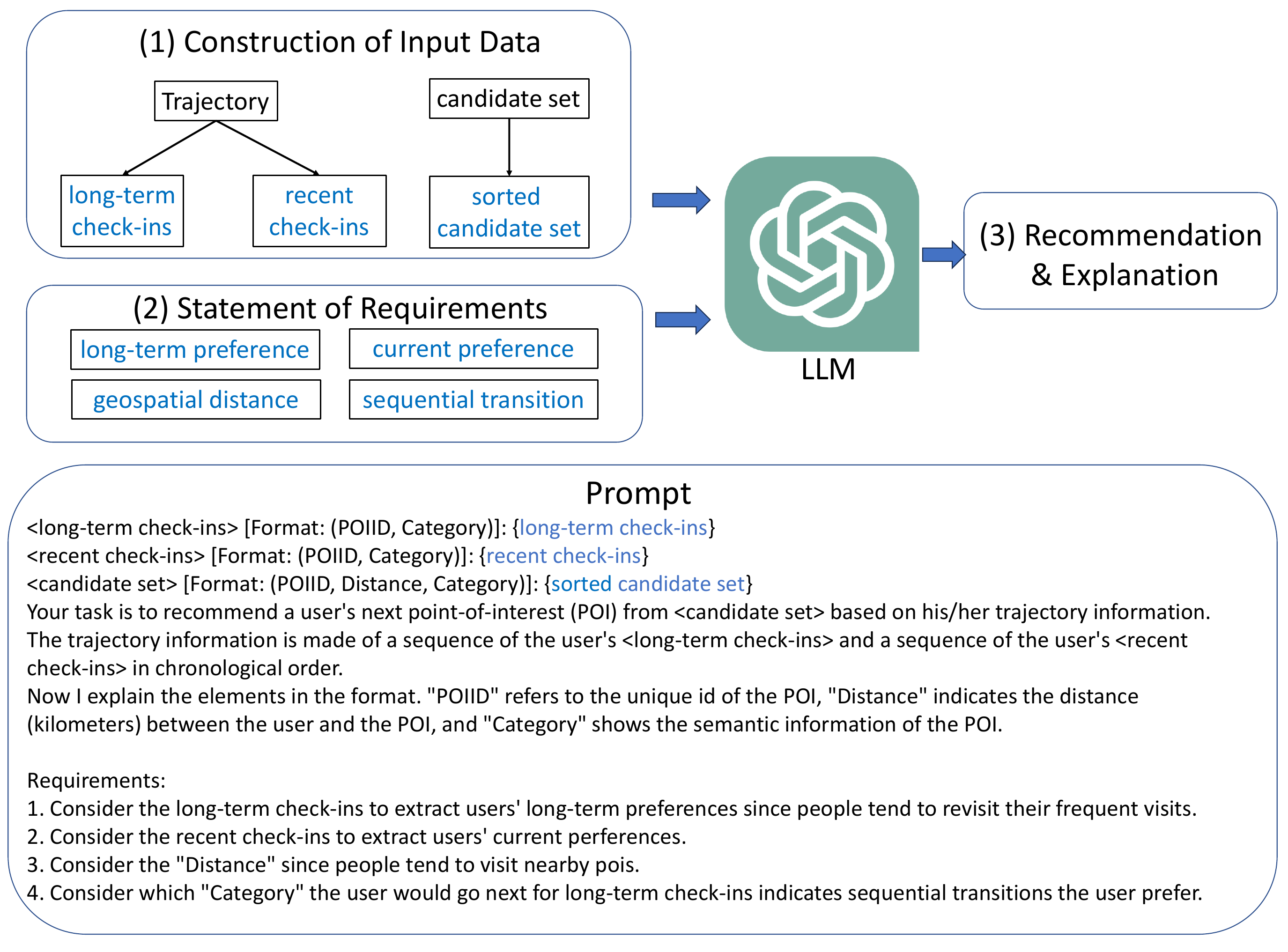}

\vspace{-1ex}
\caption{The workflow of the designed LLMmove framework and the corresponding prompts.} \label{fig:framework}
\vspace{-1ex}
\end{figure}

Our study empowers Large Language Models (LLMs) for zero-shot next Point of Interest (POI) recommendations using a multi-step prompting strategy, namely LLMmove, as depicted in Figure 3.  This LLMmove framework guides LLMs through three phases. Initially, it preprocesses data and incorporates background knowledge as input. Subsequently, it instructs the LLM to weigh four key factors: long-term and current user preferences, geographical distance, and sequential transitions. Finally, the LLM generates top-\(k\) POI recommendations with accompanying explanations.

\subsection{Construction of Input Data}

For personalized next POI recommendations, we leverage two types of user trajectory data: long-term check-ins capturing broader preferences and recent check-ins reflecting immediate interests. Additionally, we factor in the geographical distance of candidate POIs from the user's current location, recognizing its impact on travel behavior. These elements feed into the LLM, enabling it to suggest relevant and timely locations for each user.

\subsection{Statement of Requirements}

When analyzing user mobility behaviors, four key factors demand consideration. The first is long-term check-ins, offering insights into user preferences. The second, recent check-ins, mirror current contextual preferences. Thirdly, distance plays a role, as users tend to favor nearby Points of Interest (POI). Additionally, we delve into sequential transition patterns, exploring the flow between continuous categories in users' long-term check-ins.

\subsection{Recommendation and Explanation}

In this stage, we guide the LLM to generate the next POI recommendation along with reasons for the suggestion. Leveraging background knowledge, including both long-term and recent check-ins, and considering the candidate side, the LLM is instructed to incorporate the four requirements into the user movement suggestion. Given the conditions and requirements, the LLM produces recommendations and explanations.

\section{Experiments} \label{sec:exp}
\subsection{Experimental Setup}

\subsubsection{Datasets}
To assess the efficacy of leveraging LLMs for next-POI recommendation, we employ two widely-used datasets: NYC~\cite{dataset_yang2014modeling} and TKY~\cite{dataset_yang2014modeling}, following the experimental setting of ~\cite{yan2023_st_hypergraph_sigir}\footnote{https://github.com/ant-research/Spatio-Temporal-Hypergraph-Model}. Both the NYC and TKY datasets were collected from the Foursquare from April 2012 to February 2013,  constituting location-based social network check-in records. NYC denotes the check-in data in New York City, while TKY covers the check-in data from Tokyo. Each check-in record includes the user, POI, POI category, geographical coordinates, and timestamp. The check-in records are organized in chronological order, with the first 80\% serving as the training set, the subsequent 10\% as the validation set, and the last 10\% as the test set. The statistics of check-in datasets are reported in Table~\ref{table:dataset}. Here, \#Test-Traj denotes the number of trajectories in test cases. For a given test trajectory $Traj=\{l_1, l_2, ..., l_k\}$, we utilize the first $k-1$ check-ins are the recent context, and the last visited POI $l_k$ is considered as the ground truth POI. 
In each test case, the candidate set comprises the ground truth POI and 100 randomly sampled POIs, following a common practice in recommendation evaluation~\cite{WWW17_NCF_he2017neural}.


It's noteworthy that historical records in the training set are exclusively used to reflect users' long-term preferences. Instead of constructing sequential POI recommendation models, as in existing studies, we directly employ LLMs to perform zero-shot POI recommendations for users' next moves.

\begin{table}[htb!] 
\begin{center}  
\vspace{-1ex}
\caption{Statistics of two real-world check-in datasets} \label{table:dataset}
    \begin{tabular}{| p{0.9cm}| p{0.8cm} | p{0.8cm} | p{1.1cm} | p{1.2cm}| p{1.2cm}| }
    \hline
    Dataset & \#Users & \#POIs  & \#Category  & \#Checkins & \#Test-Traj \\ \hline
	NYC &  1,048  & 4,981  & 318 & 103,941 & 1,364 \\ \hline
	TKY &  2,282 & 7,833  & 290  & 405,000  & 4,610 \\ 
    \hline
    \end{tabular}
\end{center}
\end{table}

\begin{table*}[!htb]
\begin{center}
\caption{The experimental results of next-POI recommendation on real-world datasets.} \label{table:exp_results}
\vspace{-1ex}
\begin{tabular}{|p{2.1cm}|p{1.2cm}|p{1.2cm}|p{1.2cm}|p{1.2cm}|p{1.2cm}|p{1.2cm}|p{1.2cm}|p{1.2cm}|}
  \hline 
 & \multicolumn{4}{|c|}{NYC} & \multicolumn{4}{|c|}{TKY}\\
  \hline
  Methods & Acc@1 & Acc@5 & Acc@10 & MRR & Acc@1 & Acc@5 & Acc@10 & MRR\\ \hline 
Popu & 0.0500 & 0.2200 &0.2750 & 0.1168 & 0.2300 & 0.4000 & 0.5050  & 0.3148 \\ 
Dist & 0.3702 & 0.5700 &0.6195 &0.4452 & 0.2700 & 0.4850 & 0.5450  & 0.3682 \\
CZSR & 0.1600 & 0.2400 & 0.2800 & 0.1903 & 0.1300 & 0.1700 & 0.1800  & 0.1461\\ 
LLMRank & 0.0400 & 0.1150 & 0.1750 & 0.0738 & 0.0100 & 0.0850 & 0.1600  & 0.0469\\ 
ListRank& 0.1100 & 0.1700 & 0.2050 & 0.1347 & 0.1250 & 0.1500 & 0.1750  & 0.1357\\ \hline
LLMMob & 0.3600 & 0.5450 & 0.6050 & 0.4384 & 0.3350 & 0.4800 & 0.5150  & 0.3873\\ 
LLMMOB(-time) & 0.2750 & 0.5550 & 0.6500 & 0.3911 & 0.2750 & 0.4550 & 0.4900 & 0.3505\\ 
LLMMob(+Geo) & 0.3850 & 0.5850 & 0.6550 & 0.4703 & 0.3200 & 0.5500 & 0.6100  & 0.4091  \\ \hline
LLMmove & 0.5200 & 0.6100 & 0.6650 & 0.5585 & 0.4200 & 0.5800 & 0.6250  & 0.4847  \\
 \hline
\end{tabular}
\vspace{-1ex}
\end{center}
\end{table*}

\subsubsection{Evaluated Methods}
We assess the performance of the following approaches:
\begin{itemize}
    \item \textbf{Popu}: This method selects the most popular POIs, i.e., the frequently visited locations by users. 
    \item \textbf{Dist}: This approach directly chooses the nearest locations, i.e., the locations with the shortest distance.
    \item \textbf{CZSR}~\cite{chatgpt_zero_liu2023chatgpt}: It develops a set of prompts for different recommendation scenarios, and we choose the zero-shot sequential recommendation prompt as the compared baseline in this work.
    \item \textbf{LLMRank}~\cite{Zero-shot-ranker-hou2023large}: LLMRank regards the recommendation task as a ranking task, where historical interactions serve as conditions, and LLMs are instructed to rank a set of candidates. We choose the recency-focused prompting to incorporate recent check-in information.
    \item \textbf{ListRank}~\cite{list_ranking_dai2023uncovering}: It boosts the LLM’s recommendation capabilities by ranking policies. We use the list-wise ranking policy as the compared recommendation baseline. 
    
    \item \textbf{LLMMob}~\cite{Arxiv23_MOB_wang2023would}: It leverages LLMs to analyze human mobility data by considering both long-term and short-term dependencies. By incorporating the temporal information, it aims at solving the time-aware mobility prediction.
     \item \textbf{LLMMob(-Time)}~\cite{Arxiv23_MOB_wang2023would}: To make a fair comparison with other baselines, it removes the temporal information in the LLMMob framework. 
    \item \textbf{LLMMob(+Geo)}: It extends the LLMMob approach~\cite{Arxiv23_MOB_wang2023would} by additionally considering geographical influence, which is the same with the setting of LLMmove.
    \item \textbf{LLMmove}: LLMmove extracts user preferences, geographical influence, and sequential transitions, which constitute our proposed prompting strategy.    
\end{itemize}

For all the evaluated LLM-based methods, we use the \textit{gpt-3.5-turbo} as the default LLM for a fair comparison.

\subsubsection{Evaluation Metrics}

Consistent with prior studies on POI recommendation~\cite{SIGIR22_yang2022getnext, yan2023_st_hypergraph_sigir}, we employ two widely used performance evaluation metrics: Top-K accuracy rates (Acc@K) and Mean Reciprocal Rank (MRR). Acc@k assesses whether the ground truth POI is present in the Top-K recommended list, while MRR considers the ranking position of the ground truth in the sorted recommended list. 
Given the $n$ test cases, Acc@k and MRR are defined as:
\begin{equation}
\small
Acc@k = \frac{1}{n} \sum_{1}^{n} hits(rank_{gt} \leq k), \:
MRR = \frac{1}{n} \sum_{1}^{n}\frac{1}{rank_{gt}}. 
\end{equation}

The function $hits()$ is an indicator: it returns 1 if the condition is true, and 0 otherwise. $rank_{gt}$ represents the rank of the ground truth next POI in the recommended list.


\subsection{Next-POI Recommendation Results}
The empirical results for next-POI recommendation are presented in Table~\ref{table:exp_results}. The observations are as follows: (1) Popu demonstrates satisfactory performance, suggesting users' inclination towards popular places. (2) Dist shows promising results, surpassing existing sequential zero-shot recommendation methods significantly. This underscores the crucial role of geographical distance in the next POI recommendations. (3) Applying existing sequential zero-shot methods (CSR, LLMRank, ListRank) for predicting users' next movements is unfeasible, as they lack geospatial consideration. (4) LLMMob achieves relatively higher scores on both datasets, indicating its effectiveness in modeling user mobility data. By comparing the three variants of LLMMob, we can learn that both temporal and spatial information are beneficial for predicting future movements. (5) The proposed LLMMove attains the best performance, highlighting the advantages of the prompting strategy. Different from other baselines, it can make full use of geographical distance and sequential transition patterns. In particular, LLMmove outperforms the LLMMob(+Geo)by effectively leveraging spatial distance and personalized sequential patterns in user check-in behaviors. Overall, the proposed LLMmove excels in the next POI recommendation, showcasing its efficacy in harnessing LLM capabilities and learning user movement trajectories.


\begin{table}[!htb]
\vspace{-1ex}
\begin{center}
\caption{The ablation studies of LLMmove on the NYC dataset.} \label{table:ablation_results}
\vspace{-1ex}
\begin{tabular}{|p{1.5cm}|p{1.0cm}|p{1.0cm}|p{1.0cm}|p{1.0cm}|}
  \hline 
  Variants & Acc@1 & Acc@5 & Acc@10 & MRR \\ \hline 
LLMmove & 0.5200 & 0.6100 & 0.6650 & 0.5585  \\  \hline
-LP & 0.4900 & 0.5800 & 0.6200 & 0.5266  \\
-RP & 0.5100 & 0.6100 & 0.6500 & 0.5521  \\ 
-Geo & 0.4800 & 0.5750 & 0.6250 & 0.5201 \\ 
-Seq & 0.5250 & 0.5850 & 0.6400 & 0.5554 \\  \hline
\end{tabular}
\vspace{-1ex}
\end{center}
\end{table}


\begin{table}[!htb]
\vspace{-1ex}
\begin{center}
\caption{The impact of the order of candidate POI on the NYC dataset.} \label{table:ablation_results_order}
\begin{tabular}{|p{1.5cm}|p{1.0cm}|p{1.0cm}|p{1.0cm}|p{1.0cm}|}
  \hline 
  Order & Acc@1 & Acc@5 & Acc@10 & MRR \\ \hline 
(a) Dist-asc & 0.5200 & 0.6100 & 0.6650 & 0.5585  \\  \hline
(b) Dist-des & 0.2000 & 0.3000 & 0.3250 & 0.2398  \\
(c) Rand & 0.3250 & 0.4500 & 0.5150 & 0.3854 \\
(d) Freq-asc & 0.3600 & 0.4450 & 0.4950 & 0.4060  \\ 
(e) Freq-des & 0.4400 & 0.5400 & 0.6400 & 0.4900 \\ \hline
\end{tabular}
\vspace{-1ex}
\end{center}
\end{table}

\subsection{Ablation Studies}
We conduct ablation studies to examine the impacts of different factors: long-term preference (LP), recent preference (RP), geographical influence (Geo), and sequential transition (Seq). The empirical results are presented in Table~\ref{table:ablation_results}, yielding the following observations: 
(1) The considerable performance gap between LLMmove-LP and LLMmove indicates the importance of long-term preference.
(2) LLMmove-RP shows results close to LLMmove, indicating that current preference might not be significantly influential in LLMs. Despite the acknowledged relevance of a user's current interest in POI recommendations, the LLMs may struggle to fully utilize this factor in zero-shot scenarios, lacking collaborative information from other users.
(3) LLMmove-Geo exhibits the lowest performance, underscoring the essential role of geographical distance in next POI recommendations.
(4) LLMmove-Seq attains results similar to LLMmove. Although it shows slightly better results on Acc@1, its performance is inferior on Acc@5, Acc@10, and MRR. The relatively lower scores of Acc@5 and Acc@10 imply that, without explicitly considering this requirement, LLM may overlook potential POIs associated with sequential transitions.



Moreover, we made an interesting observation: the order of the candidate set significantly affects the recommendation performance. To explore this impact, we compared several ways to present the candidate POIs: (a) sorted by their distances in ascending order (Dist-asc); (b) sorted by their distances in descending order (Dist-des); (c) random sort (Rand); (d) sorted by their category frequencies in ascending order (Freq-asc); (e) sorted by their category frequencies in descending order (Freq-des).   
The results are reported in Table~\ref{table:ablation_results_order}. Notably, presenting POIs with higher probabilities (e.g., shorter distance or higher popularity) at the beginning of the candidate set leads to remarkable improvement. For example, the performance of Dist-asc is significantly better than Rand and Dist-des. This improvement could be attributed to the LLM's limited capability to handle a large number of candidates. As an empirical choice, we adopt the Dist-asc ordering to present POI candidates in our work.

\begin{table}[!htb]
\begin{center}
\vspace{-2ex}
\caption{Cases of reasons. The \textcolor{red}{red} text highlights the requirements, and the ground truth POI is marked in \textcolor{blue}{BLUE}.}
\vspace{-2ex}
\label{table:cases}
\begin{tabular}{p{8.5cm}}
\hline
\footnotesize
    {\centering Case 1 (\textbf{Correct})\\}
    
    ``recommendation'': [``4975'', ``1449'', ``3864'', ``2713'', ``2525'', ``3513'', ``356'', ``2836'', ``3245'', ``4321'']. \\
    ``reason'': ``\textcolor{red}{Based on the user's long-term check-ins}, the user frequently visits POI `1449' (Gym / Fitness Center) and `2836' (Neighborhood). Therefore, it is likely that the user will visit these places again. Additionally, \textcolor{red}{considering the recent check-ins}, the user has visited POI `1449' (Gym / Fitness Center) and `3864' (Neighborhood) recently, indicating a current preference for these places. Furthermore, \textcolor{red}{the user tends to visit nearby pois}, so POI `4975' (Home (private)) and `2713' (Subway) are also recommended. Finally, \textcolor{red}{considering the sequential transitions} in the long-term check-ins, the user may visit POI `3245' (Elementary School) after visiting `2836' (Neighborhood).''\\
    ``groundtruth'': \textcolor{blue}{4975}\\
    \hline
    {\centering Case 2 (\textbf{Correct})\\}
    ``recommendation'': [``1395'', ``1494'', ``646'', ``327'', ``4853'', ``2442'', ``2932'', ``2222'', ``212'', ``1043''].\\
    ``reason'': ``\textcolor{red}{Based on the user's long-term check-ins}, the user frequently visits 'Plaza' (POIID: 1494) and 'Performing Arts Venue' (POIID: 1709). Therefore, the user is likely to prefer places in the `Plaza' category. Additionally, \textcolor{red}{considering the recent check-ins}, the user's most recent check-in was at a 'Park' (POIID: 2379). \textcolor{red}{Taking into account the distance}, the closest POI in the candidate set is `Law School' (POIID: 1395). Therefore, the recommended POIs are the ten most probable places in the candidate set that match the user's preferences and are closest in distance.''\\
    ``groundtruth'': \textcolor{blue}{1494}\\
    \hline
    
    {\centering Case 3 (\textbf{Incorrect})\\}
    
    ``recommendation'': [``4882'', ``3252'', ``1986'', ``4604'', ``1297'', ``4676'', ``3081'', ``1122'', ``1232'', ``1304''].\\
    ``reason'': ``\textcolor{red}{Based on the user's long-term check-ins}, the user frequently visits POIIDs `1823' (Bar), `2279' (Deli / Bodega), `3481' (Home (private)) and `1823' (Bar). Therefore, the user is likely to prefer places in the `Bar', `Deli / Bodega', and `Home (private)' categories. \textcolor{red}{Considering the recent check-ins}, the user has visited POIIDs `242' (Bookstore) and `1048' (Toy / Game Store). \textcolor{red}{Based on the distance}, the user is likely to prefer nearby places. Therefore, the recommended POIIDs are the ten most probable places in the candidate set that match the user's preferences and are nearby.''\\
    ``groundtruth'': \textcolor{blue}{356}
\\\hline
\end{tabular}
\vspace{-2ex}
\end{center}
\end{table}


\subsection{Case Studies}
To develop an intuitive understanding of LLMmove's reasoning ability and interpretability, we select and analyze some example results from the NYC dataset, as shown in Table~\ref{table:cases}. Case 1 and case 2 demonstrate that LLMmove can deliver accurate recommendations accompanied by plausible reasons. These reasons align with the given requirements, showcasing the model's ability to reflect the specified factors for recommendations. However, LLMMove fails to give reasonable recommendations sometimes. From Case 3, we can observe that LLMMove points out the user's frequently visited POI '1823' (Bar) twice, which may be attributed to limited categories in the data. Additionally, when faced with insufficient sequential transitions, LLM may struggle to make accurate predictions.

\section{Conclusion}


In this study, we concentrate on harnessing the capabilities of Large Language Models (LLMs) for the zero-shot next POI recommendation task. Our approach considers both the users' long-term preferences and current preferences, as well as geographic spatial distance and sequential transitions in user mobility behaviors. To integrate these factors, we introduce a novel prompt strategy aimed at generating top-K subsequent POI recommendations along with the rationale for suggestions. Extensive experiments conducted on two real-world datasets demonstrate that our proposed method significantly outperforms existing LLM-based baselines, showcasing its effectiveness in the next POI recommendation. However, our findings highlight potential challenges in spatial reasoning and understanding geographical information, emphasizing the need for future research to enhance the performance of generative POI recommendations.


\bibliographystyle{IEEEtran}
\bibliography{ref}
\end{document}